\documentclass[nofootinbib,preprint,aps,draft,pre,superscriptaddress,citeautoscript,floatfix]{revtex4-2}
\setcitestyle{super}
\usepackage{graphics}
\usepackage{graphicx}
\usepackage{mathrsfs}
\usepackage{amsmath}
\usepackage{bm}
\usepackage{color}
\usepackage{float}
\usepackage[marginal]{footmisc}

\renewcommand\keywords[1]{\textbf{Keywords}: #1}
\begin{document}
%%%%%%%%%%%%%%%%%%%%%%%%%%%%%%%%%

\title{The process-directed self-assembly of block copolymer particles}
\author{Yanyan Zhu}
\affiliation{Center of Soft Matter Physics and its Applications, School of Physics, Beihang University, Beijing 100191, China}
\author{Changhang Huang}
\affiliation{Center of Soft Matter Physics and its Applications, School of Physics, Beihang University, Beijing 100191, China}
\author{Liangshun Zhang}
\email{zhangls@ecust.edu.cn}
\email{andelman@tauex.tau.ac.il}
\email{manxk@buaa.edu.cn}
\affiliation{Shanghai Key Laboratory of Advanced Polymeric Materials, Key Laboratory for Ultrafine Materials of Ministry of Education, School of Materials Science and Engineering, East China University of Science and Technology, Shanghai 200237, China}
\author{David Andelman}
\email{zhangls@ecust.edu.cn}
\email{andelman@tauex.tau.ac.il}
\email{manxk@buaa.edu.cn}
\affiliation{School of Physics and Astronomy, Tel Aviv University, Ramat Aviv 69978, Tel Aviv, Israel}
\author{Xingkun Man}
\email{zhangls@ecust.edu.cn}
\email{andelman@tauex.tau.ac.il}
\email{manxk@buaa.edu.cn}
\affiliation{Center of Soft Matter Physics and its Applications, School of Physics, Beihang University, Beijing 100191, China}
\affiliation{Peng Huanwu Collaborative Center for Research and Education, Beihang University, Beijing 100191, China}

%%%%%%%%%%%%%%%%%%%%%%
\begin{abstract}
%%%%%%%%%%%%%%%%%%%%%%

We explore the kinetic paths of structural evolution and formation of block copolymer (BCP) particles using dynamic self-consistent field theory (DSCFT). We show that the process-directed self-assembly of BCP immersed in a poor solvent leads to the formation of striped ellipsoids, onion-like particles and double-spiral lamellar particles. The theory predicts a reversible path of shape transition between onion-like particles and striped ellipsoidal ones by regulating the temperature (related to the Flory-Huggins parameter between the two components of BCP, $\chi_{\rm{AB}}$) and the selectivity of solvent towards one of the two BCP components. Furthermore, a kinetic path of shape transition from onion-like particles to double-spiral lamellar particles, and then back to onion-like particles is demonstrated. By investigating the inner-structural evolution of a BCP particle, we identify that changing the intermediate bi-continuous structure into a layered one is crucial for the formation of striped ellipsoidal particles. Another interesting finding is that the formation of onion-like particles is characterized by a two-stage microphase separation. The first is induced by the solvent preference, and the second is controlled by the thermodynamics. Our findings lead to an effective way of tailoring nanostructure of BCP particles for various industrial applications.


\end{abstract}

\maketitle

%%%%%%%%%%%%%%%%%%%%%%%%%%%%%%%%%%%%%%%%%%%%%%%%%%%%%%%
\section{Introduction}
%%%%%%%%%%%%%%%%%%%%%%%%%%%%%%%%%%%%%%%%%%%%%%%%%%%%%%%

Block copolymer (BCP) particles have attracted considerable scientific interest due to their numerous applications for sensors, smart coating, drug delivery and more~\cite{Choi14,Shin18,Venkataraman11}. The particle rheological and optical properties highly depend on their apparent shape and inner nanostructure~\cite{Yunker11,Forster11,Yoo10}. Nevertheless, the precise tailoring of BCP particle structure and apparent shape on the nanoscale remains an outstanding challenge~\cite{Ku18,Shi13,Yan18}.

Emulsions are often used to fabricate BCP particles~\cite{Staff13,Jang13,Shin15}. In such a process, BCPs are first dissolved in a good solvent, such as toluene or chloroform. Then, the solution is emulsified in an aqueous phase and forms a dispersion of emulsion droplets in water. Surfactants are usually added to the aqueous phase in order to stabilize the droplets. At first, the emulsion droplets contain both BCPs and the organic solvents, where the BCPs are in a disordered state due to their low concentration. As the organic solvent evaporates, the droplets deswell and the BCP concentration increases, causing a microphase separation of BCP within each of the emulsion droplets~\cite{Wyman11,Kim15,Shin20}. Eventually, an aqueous suspension of BCP particles is obtained, where the BCP particles have diverse nanostructures, including striped ellipsoidal, onion-like, oblate as well as patchy particles~\cite{Zhu22,Deng13,Jeon08,Ku15,Zhu19}.

As mentioned above, once the particles have been obtained, solvent vapor annealing (SVA) and surfactant are commonly used to regulate both the apparent shape and inner structure of these particles~\cite{Deng14,Shin19,Navarro22,Navarro221,Tan22}. In more details, the suspension is first put into a close chamber containing organic solvent vapor. The solvent dissolves, penetrates into the BCP-rich region and swells the particles. Then, in a second stage the swelled particles are put into another chamber without the organic solvent vapor, leading to solvent evaporation, and deswelling of the BCP particles. The main solvent effect during the SVA process is to enhance the mobility of the polymer chains, facilitating their microphase separation. Surfactant is added in order to control the interfacial tension between the BCP particles and their surrounding environment. When the surfactant has a preference towards one of BCP components, usually onion-like particles are formed. On the other hand, when the surfactant is neutral towards the BCP, striped ellipsoidal particles are formed~\cite{Klinger14,Deng132,Ku182}. In recent experiments, it was shown that the surfactant selectivity can be controlled by using stimuli-responsive surfactants that are sensitive to pH, temperature and light~\cite{Lee17,Lee19,Lee192,Kim21,Kim22}. For example, the azobenzene-containing surfactant is photoswitchable. By irradiating visible or UV light on the particles, a reversible shape transition occurs between onion-like and ellipsoidal shapes~\cite{Hu21,Kwon21}.

Using SVA with added surfactant to produce BCP particles is a complex multistage process involving temperature, evaporation rate, surface preference, and shear stress~\cite{Wang22,Shin17,Li10}. In such a complicated fabrication process, the particles can easily be trapped in metastable states, resulting in a final structure that is not the desired one. However, there are only a few experimental and theoretical studies addressing the kinetic path of the BCP particle formation, and the process-directed self-assembly of BCP particles remains poorly understood~\cite{Muller09,Sun17,Abetz19,Ma18,Cai22}. In this work, we directly address this problem by investigating the coupled shape and inner-structural evolution of BCP particles in a thermal annealing (TA) process. Thermal annealing has been preliminary used in the past to induce shape transition of BCP particles~\cite{Higuchi10,Higuchi13,Avalos18}. Analyzing the kinetic path of particle formation in the TA process is valuable for understanding and optimizing current SVA experiments that fabricate BCP particles where temperature plays a key factor. Moreover, TA has the advantage of offering a simple way of tailoring the structure of BCP particles because it involves fewer factors as compared to SVA.

Compared to numerous experimental studies, there have been relatively few theoretical calculations and simulations regarding the formation of BCP particles. Li et al utilized self-consistent field theory (SCFT) to investigate the impact of particle size on the apparent shape of particles~\cite{Hsu13,Xia14}, while Sevink et al employed dynamic self-consistent field theory (DSCFT) to study the kinetic path of BCP particle formation under given conditions~\cite{Fraaiji03,Sevink05}. Additionally, several theoretical studies have focused on the self-assembly of block copolymers under soft confinement, with a particular emphasis on vesicle formation~\cite{He01,Uneyama07,Chi11,Wang11}. Recent reviews have highlighted the strong dependence of BCP system equilibrium states on the process-directed self-assembly behavior of BCPs~\cite{Li16,Muller20}. However, this behavior has yet to be thoroughly studied in relation to the formation of BCP particles. To address this problem, we use dynamic self-consistent field theory (DSCFT) to investigate the structural evolution of BCP particles as a function of the Flory-Huggins interaction parameter between the AB blocks. Herein, in section 2 we present the reversible transition between striped ellipsoidal particles and onion-like ones induced by thermal annealing, followed by the formation process of these ordered particles. We summarize the conclusions in section 3. Finally, we introduce our model that is based on the DSCFT in section 4.

%%%%%%%%%%%%%%%%%%%%%%%%%%%%%%
\section{Results and Discussions}
%%%%%%%%%%%%%%%%%%%%%%%%%%%%%%

We use DSCFT to investigate the kinetic path of reversible shape transition of a BCP particle embedded in a poor solvent. We thoroughly examine the structural evolution of BCP particles as their inner structure changing from ($\it{i}$) disordered to striped ellipsoid; ($\it{ii}$) disordered to onion-like; and, ($\it{iii}$) striped ellipsoid to onion-like. The aim of our study is to present a comprehensive understanding of such kinetic paths of BCP particles, and show how their inner structure evolves in response to changes of experimental conditions.

We consider two solvents with distinct selectivity, which is determined by surfactants and remains independent of temperature. The first is a neutral solvent with respect to the two blocks, satisfying $N\chi_{\rm{AC}}=N\chi_{\rm{BC}}=15$, while the second is a selective solvent preferring the B component, $N\chi_{\rm{AC}}=21$ and $N\chi_{\rm{BC}}=9$. The heating process in the thermal annealing (TA) is achieved by decreasing $N\chi_{\rm{AB}}$, while the cooling process mimics the increase of $N\chi_{\rm{AB}}\sim 1/T$.

%%%%%%%%%%%%%%%%%%%%%%%%%%%%%%%%%%%%%%%%%%%%%%%%%%
\subsection{Kinetic path of reversible transition}
%%%%%%%%%%%%%%%%%%%%%%%%%%%%%%%%%%%%%%%%%%%%%%%%%%

%fig1
%%%%%%%%%%%%%%%%%%%%%%%%%%%%%%%%%%%%%%%%%%%%%%%%%%%%%%%%%%%%%%%%%%%%%%%%%%%%%%%%%%%%%%%%%%%%
\begin{figure*}[h!t]
{\includegraphics[width=1.0\textwidth,draft=false]{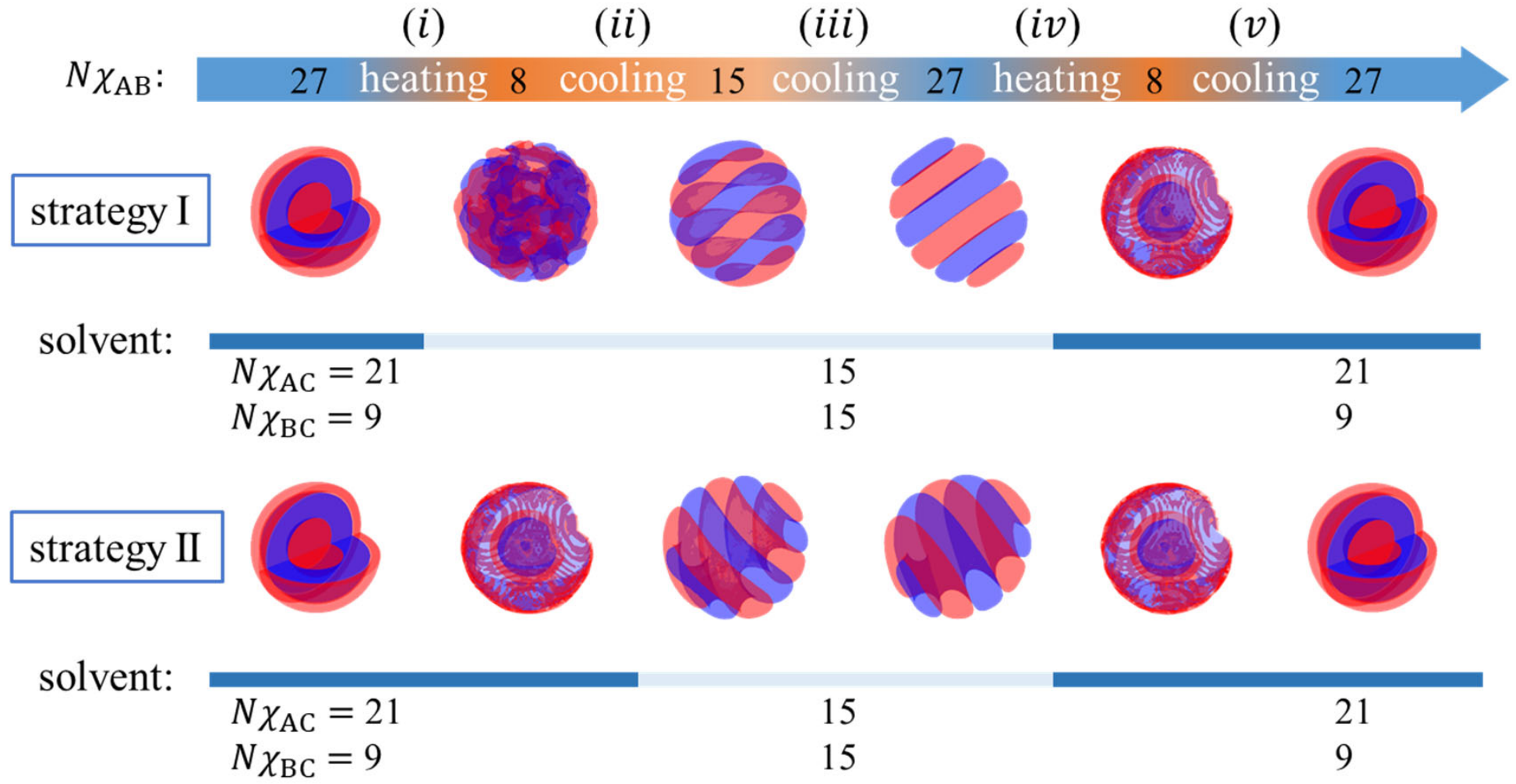}}
\caption{
\textsf{Kinetic paths of the shape transition by varying $N\chi_{\rm{AB}}$. Strategy I shows a path of a reversible transition between onion-like and striped ellipsoidal particle, while strategy II shows a path of a reversible transition between onion-like and double-spiral lamellar particle. We mimic the heating process by decreasing $N\chi_{\rm{AB}}$, while the increase of $N\chi_{\rm{AB}}$ corresponds to a cooling process. The deep blue color in the solvent color bar represents selective solvent environment, chosen as $N\chi_{\rm{AC}}=21$, $N\chi_{\rm{BC}}=9$, and the light blue represents neutral solvent environment, by setting $N\chi_{\rm{AC}}=N\chi_{\rm{BC}}=15$. Inside the particle, the A-rich domains are colored in blue and B-rich are in red.}}
\label{fig1}
\end{figure*}
%%%%%%%%%%%%%%%%%%%%%%%%%%%%%%%%%%%%%%%%%%%%%%%%%%%%%%%%%%%%%%%%%%%%%%%%%%%%%%%%%%%%%%%%%%%%

Figure~\ref{fig1} shows the shape transition of a BCP particle undergoing a TA process. We show two paths (strategy I and strategy II) to indicate the process-directed self-assembly of BCP particles. Using DSCFT, the initial state for the two kinetic paths (I and II) is an onion-like particle embedded in a selective solvent environment, prepared by setting $N\chi_{\rm{AC}}=21$ and $N\chi_{\rm{BC}}=9$, and $N\chi_{\rm{AB}}=27$. The initial condition for the onion-like particle has the value of $\phi_{\rm{A}}$ as an arbitrary number between $0$ and $1$, $\phi_{\rm{B}}=1-\phi_{\rm{A}}$ and $\phi_{\rm{C}}=0$ in a preassigned BCP-rich domain, while $\phi_{\rm{A}}=\phi_{\rm{B}}=0$ and $\phi_{\rm{C}}=1$ in the surrounding solvent-rich domain.

Strategy I (the top panel in figure~\ref{fig1}) is a kinetic path of a reversible transition between onion-like and striped ellipsoidal particle. This path includes five steps that are indicated in the figure: ($\it{i}$) heating the onion-like particle in a neutral solvent environment to a disordered state. Then, a two-step cooling includes: ($\it{ii}$) cooling the particle into its weak-segregation regime; and ($\it{iii}$) further cooling into the strong-segregation regime, resulting in a perfect striped ellipsoidal particle. Subsequently, ($\it{iv}$) heating the particle in a selective solvent environment into its weak-ordered state. Finally, ($\it{v}$) cooling into the strong-segregation regime will result in a return to the original onion-like particle.

Strategy II (the bottom panel in figure~\ref{fig1}) is a kinetic path of a transition from an onion-like particle to a double-spiral lamellar particle, and then back to an onion-like particle. This strategy also includes five steps: ($\it{i}$) heating the onion-like particle in a selective solvent environment; then, a two-step cooling in a neutral solvent environment (steps ($\it{ii}$) and ($\it{iii}$)), results in a double-spiral lamellar particle~\cite{Dai21,Yu07,Sheng15}. Further employing the same steps ($\it{iv}$) and ($\it{v}$) as in strategy I, the original onion-like particle is obtained. The only difference between the two strategies is step ($\it{i}$). In strategy I, the heating process of step ($\it{i}$) is done in a neutral solvent environment, while in strategy II, it is done in a selective solvent environment. Our results suggest that initially heating in $\it{neutral}$ environment leads to the transition from onion-like to striped ellipsoidal particle, while a $\it{selective}$ environment leads to the transition from onion-like to double-spiral lamellar particle. It is worth noting that each structure showed in figure~\ref{fig1} represents a (meta)stable state given the specific set of parameters, as the temporal evolution of the free energy is stabilized after a sufficient number of iterations. Another point to note is that the free-energy values obtained from different steps of strategies I and II cannot be compared directly due to the significant changes in the parameter sets applied in each of the steps.

In addition, figure~\ref{fig1} shows onion-like particles can be formed from both striped ellipsoidal particle and double-spiral lamellar particle (steps ($\it{iv}$) and ($\it{v}$) of strategies I and II). This result indicates that the formation of onion-like particle has weak dependence on the kinetic path. In other words, it is easier to obtain an onion-like particle than striped ellipsoidal ones. To understand the kinetic path effects on the formation of BCP particles, we analyze the structural evolution of nanostructured particles in both strategies, as discussed in the next subsection.

%%%%%%%%%%%%%%%%%%%%%%%%%%%%%%%%%%%%%%%%%%%%%%%%%%
\subsection{Formation process of striped ellipsoidal particle}
%%%%%%%%%%%%%%%%%%%%%%%%%%%%%%%%%%%%%%%%%%%%%%%%%%

%fig2
%%%%%%%%%%%%%%%%%%%%%%%%%%%%%%%%%%%%%%%%%%%%%%%%%%%%%%%%%%%%%%%%%%%%%%%%%%%%%%%%%%%%%%%%%%%%
\begin{figure*}[h!t]
{\includegraphics[width=0.95\textwidth,draft=false]{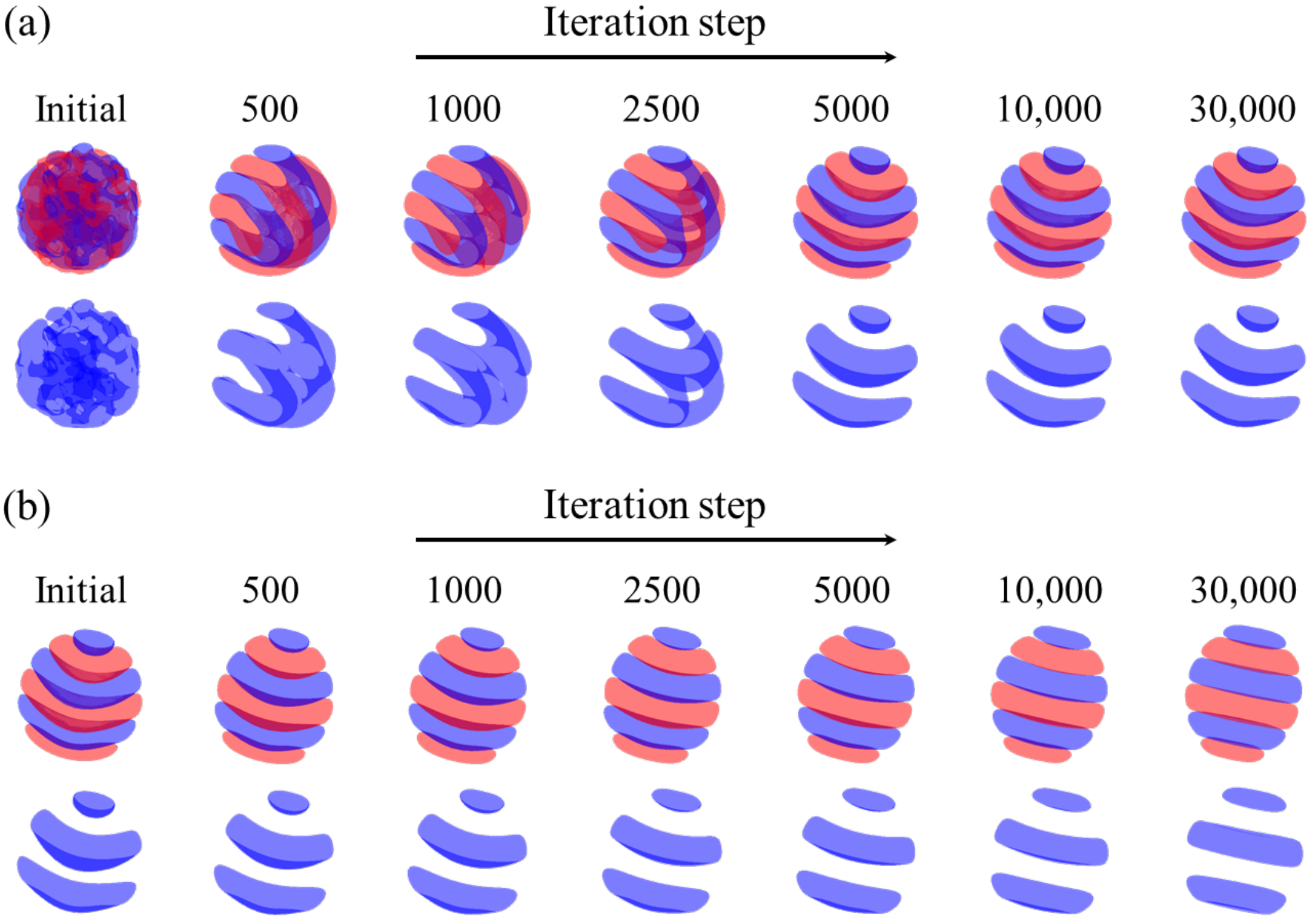}}
\caption{
\textsf{The evolution of inner structure during the formation process of a striped ellipsoidal particle. (a)~A disordered particle changes to a quasi-ellipsoidal particle upon varying $N\chi_{\rm{AB}}$ from $8$ to $15$. (b)~A quasi-ellipsoidal particle changes to a striped ellipsoidal particle upon varying $N\chi_{\rm{AB}}$ from $15$ to $27$. In both (a) and (b) the particle is embedded in a neutral solvent environment of $N\chi_{\rm{AC}}=N\chi_{\rm{BC}}=15$. The first and third rows are isosurface plots, where A-rich domains are colored in blue and B-rich domains are in red. The second and fourth rows show isosurfaces of the A domains.
}}
\label{fig2}
\end{figure*}
%%%%%%%%%%%%%%%%%%%%%%%%%%%%%%%%%%%%%%%%%%%%%%%%%%%%%%%%%%%%%%%%%%%%%%%%%%%%%%%%%%%%%%%%%%%%

Figure~\ref{fig2} shows the evolution of the inner structure of BCP particles, changing from disordered state to striped ellipsoidal state during the two-step cooling (steps ($\it{ii}$) and ($\it{iii}$) of strategy I) in a neutral solvent environment ($N\chi_{\rm{AC}}=N\chi_{\rm{BC}}=15$). The starting point is a disordered BCP particle obtained in step ($\it{i}$) of strategy I. Figure~\ref{fig2}a shows the structural evolution for the first cooling process, where the temperature is reduced from above ODT to weak-segregation regime by varying $N\chi_{\rm{AB}}$ from $8$ to $15$. Our results indicate that the nanostructure inside the particle first evolves from a disordered state to a bi-continuous one. The bi-continuous particle is formed of inner A component layers (or B component) that are connected. Then, after $\sim$2,500 numerical iteration steps, the bi-continuous state is destroyed, and a quasi-ellipsoidal particle with inner structure of curved layers is formed, as shown by isosurface plots of A-domain in the second row of figure~\ref{fig2}a. Such quasi-ellipsoidal particles have been observed in experiments~\cite{Lee192,Hu21} during the shape transition of BCP particles. Figure~\ref{fig2}b shows the second cooling process that is applied to the quasi-ellipsoidal particle by varying $N\chi_{\rm{AB}}$ from $15$ to $27$. We use $N\chi_{\rm{AB}}=27$ to be in the strong-segregation regime. The figure shows how the curved layers first become straight, and then perfect striped ellipsoidal particle is formed.

%fig3
%%%%%%%%%%%%%%%%%%%%%%%%%%%%%%%%%%%%%%%%
\begin{figure*}[h!t]
{\includegraphics[width=0.95\textwidth,draft=false]{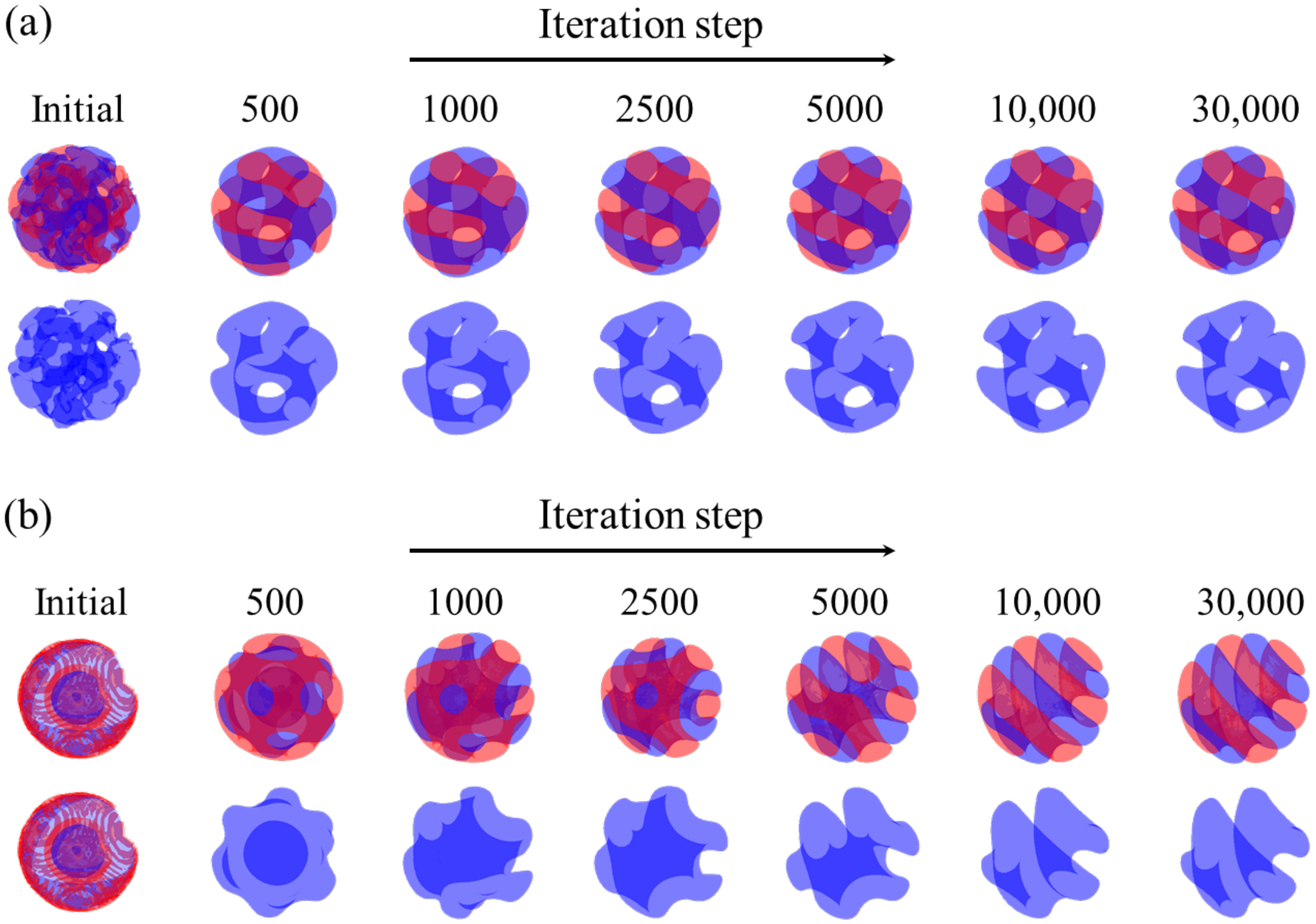}}
\caption{
\textsf{(a)~Structural evolution processes starting with a disordered particle and cooling it from $N\chi_{\rm{AB}}=8$ directly to $N\chi_{\rm{AB}}=27$. This results in a bi-continuous particle. (b)~Starting with a weak-ordered particle and cooling it from $N\chi_{\rm{AB}}=8$ to $N\chi_{\rm{AB}}=15$. This results in a double-spiral lamellar particle. In both (a) and (b) the solvent is neutral, $N\chi_{\rm{AC}}=N\chi_{\rm{BC}}=15$. The A-rich domains are colored in blue and B-rich are in red. In the second and fourth row, only the A-rich domains are shown, for clarify purposes.}}
\label{fig3}
\end{figure*}
%%%%%%%%%%%%%%%%%%%%%%%%%%%%%%%%%%%%%%%%

We would like to emphasize that the two-step cooling as well as the initial disordered particle state are crucial for the formation of the desired striped ellipsoidal particle. Figure~\ref{fig3}a shows that striped ellipsoidal particle cannot be obtained by a $\it{one}$-$\it {step}$ cooling. Starting with the same initial disordered particle as used in figure~\ref{fig2}a, the formed particle (figure~\ref{fig3}a) has an inner bi-continuous structure instead of layered structure, upon varying $N\chi_{\rm{AB}}$ from $8$ to $27$ directly ($\it{one}$-$\it {step}$ cooling). This means that $\it{one}$-$\it {step}$ cooling is not sufficient to destroy the bi-continuous structure formed in the early stages of temporal evolution.

Furthermore, figure~\ref{fig3}b shows that striped ellipsoidal particles cannot be formed also when the initial particle is changed to the one obtained in step ($\it{i}$) of strategy II. It is important to note that since step ($\it{i}$) of strategy II is a heating process in selective solvent environment, it will result in a particle with weak ordering. Such weak ordering is induced by a solvent with a selective preference to one of the BCP components. Figure~\ref{fig3}b shows the structural evolution of this weak-ordered particle during a cooling process by varying $N\chi_{\rm{AB}}$ from $8$ to $15$ in neutral solvent environment. Isosurface plots in figure~\ref{fig3}b show that a double-spiral lamellar particle is formed after 30,000 numerical iteration steps. Such a particle retains its double-spiral structure even after further cooling (varying $N\chi_{\rm{AB}}$ from $15$ to $27$), as shown in step ($\it{iii}$) of strategy II. Therefore, one of our main conclusions is that in order to obtain striped ellipsoidal particle, it is necessary to employ a kinetic path that destroys the intermediate bi-continuous structure. Otherwise, either bi-continuous particle or double-spiral lamellar particle will be formed.

%fig4
%%%%%%%%%%%%%%%%%%%%%%%%%%%%%%%%%%%%%%%%
\begin{figure*}[h!t]
{\includegraphics[width=0.5\textwidth,draft=false]{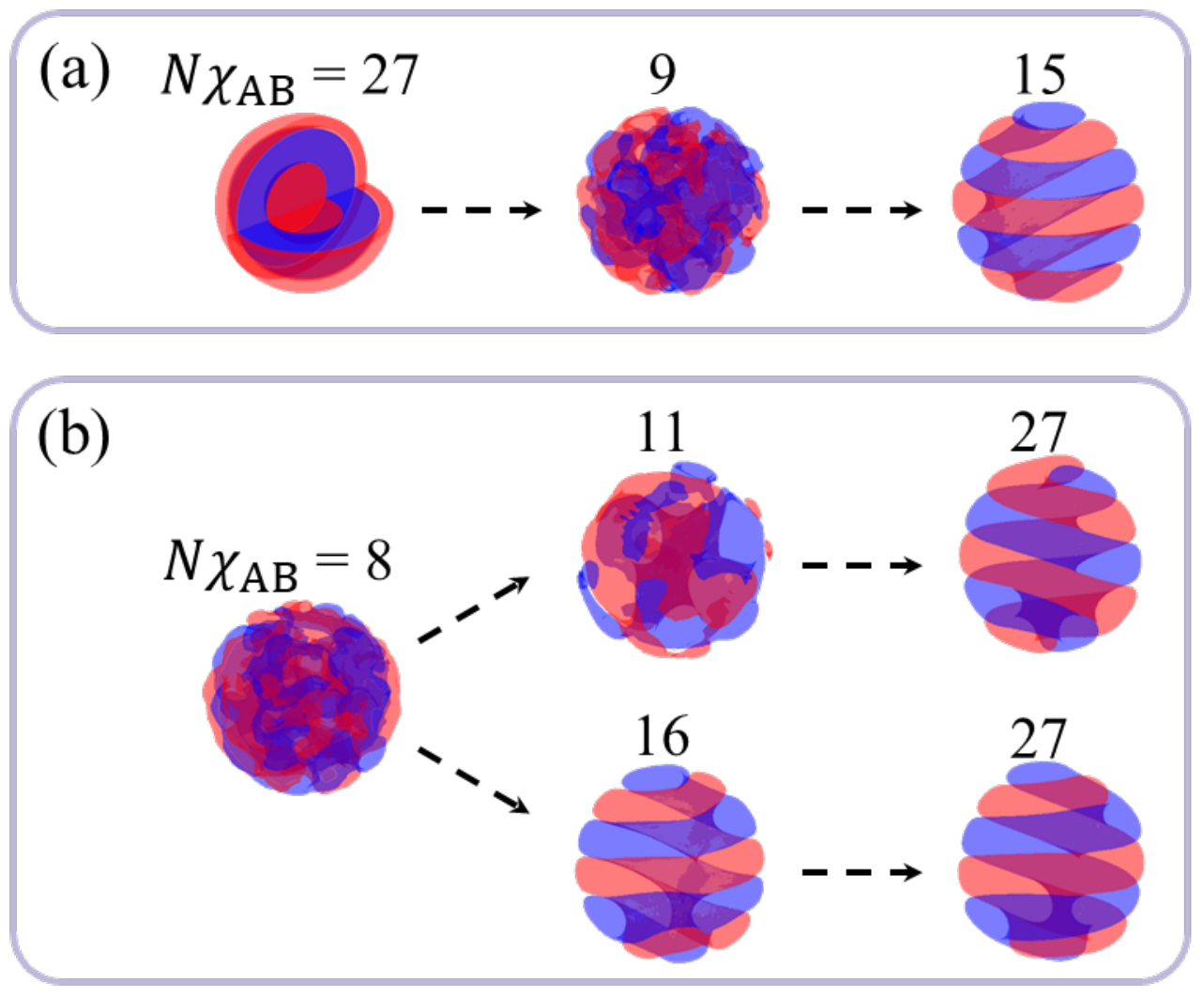}}
\caption{
\textsf{Examples for the choice of $N\chi_{\rm{AB}}$ in steps ($\it{i}$) and ($\it{ii}$) (e.g., see figure~\ref{fig1}) for the transition from onion-like to striped ellipsoidal particle. (a)~Heating an onion-like particle to $N\chi_{\rm{AB}}=9$ in step ($\it{i}$), and then cooling it to $N\chi_{\rm{AB}}=15$ in step ($\it{ii}$) results in a double-spiral lamellar particle. (b)~In the two-step cooling process, cooling a disordered particle to either $N\chi_{\rm{AB}}=11$ or $N\chi_{\rm{AB}}=16$ in step ($\it{ii}$), results in a double-spiral lamellar particle instead of a striped ellipsoid after further cooling to $N\chi_{\rm{AB}}=27$.
}}
\label{fig4}
\end{figure*}
%%%%%%%%%%%%%%%%%%%%%%%%%%%%%%%%%%%%%%%%

Another important issue for the formation of striped ellipsoidal particles is the choice of $N\chi_{\rm{AB}}$ of steps ($\it{i}$) and ($\it{ii}$) in the thermal annealing process of strategy I. The $N\chi_{\rm{AB}}$ value has to be less than $8$ ($N\chi_{\rm{AB}} \le 8$) in step ($\it{i}$). For example, figure~\ref{fig4}a shows a kinetic path where an onion-like particle is first heated to $N\chi_{\rm{AB}}=9$ in step ($\it{i}$) and then cooled to $N\chi_{\rm{AB}}=15$ in step ($\it{ii}$). Interestingly, a double-spiral lamellar particle instead of a striped ellipsoidal one is obtained. We also check that by setting $N\chi_{\rm{AB}}>9$ in step ($\it{i}$) all final particles are double-spiral (not shown in the paper). 

Our calculations demonstrate that the value of $N\chi_{\rm{AB}}$ has to be in the range between $12$ and $15$ ($N\chi_{\rm{AB}} \in [12,15]$) in step ($\it{ii}$). For example, figure~\ref{fig4}b shows a disordered particle that is cooled in step ($\it{ii}$) to either $N\chi_{\rm{AB}}=11$ or $N\chi_{\rm{AB}}=16$. Then, a further cooling to $N\chi_{\rm{AB}}=27$ is applied in both cases, leading to the formation of a double-spiral lamellar particle. Other values of $N\chi_{\rm{AB}}$ that are outside the $(12,15)$ range in step ($\it{ii}$) were also checked, and the particles always converge to double-spiral ones (not shown in the paper).

%%%%%%%%%%%%%%%%%%%%%%%%%%%%%%%%%%%%%%%%%%%%%%%%%%
\subsection{Formation process of onion-like particle}
%%%%%%%%%%%%%%%%%%%%%%%%%%%%%%%%%%%%%%%%%%%%%%%%%%

%fig5
%%%%%%%%%%%%%%%%%%%%%%%%%%%%%%%%%%%%%%%%
\begin{figure*}[h!t]
{\includegraphics[width=1.0\textwidth,draft=false]{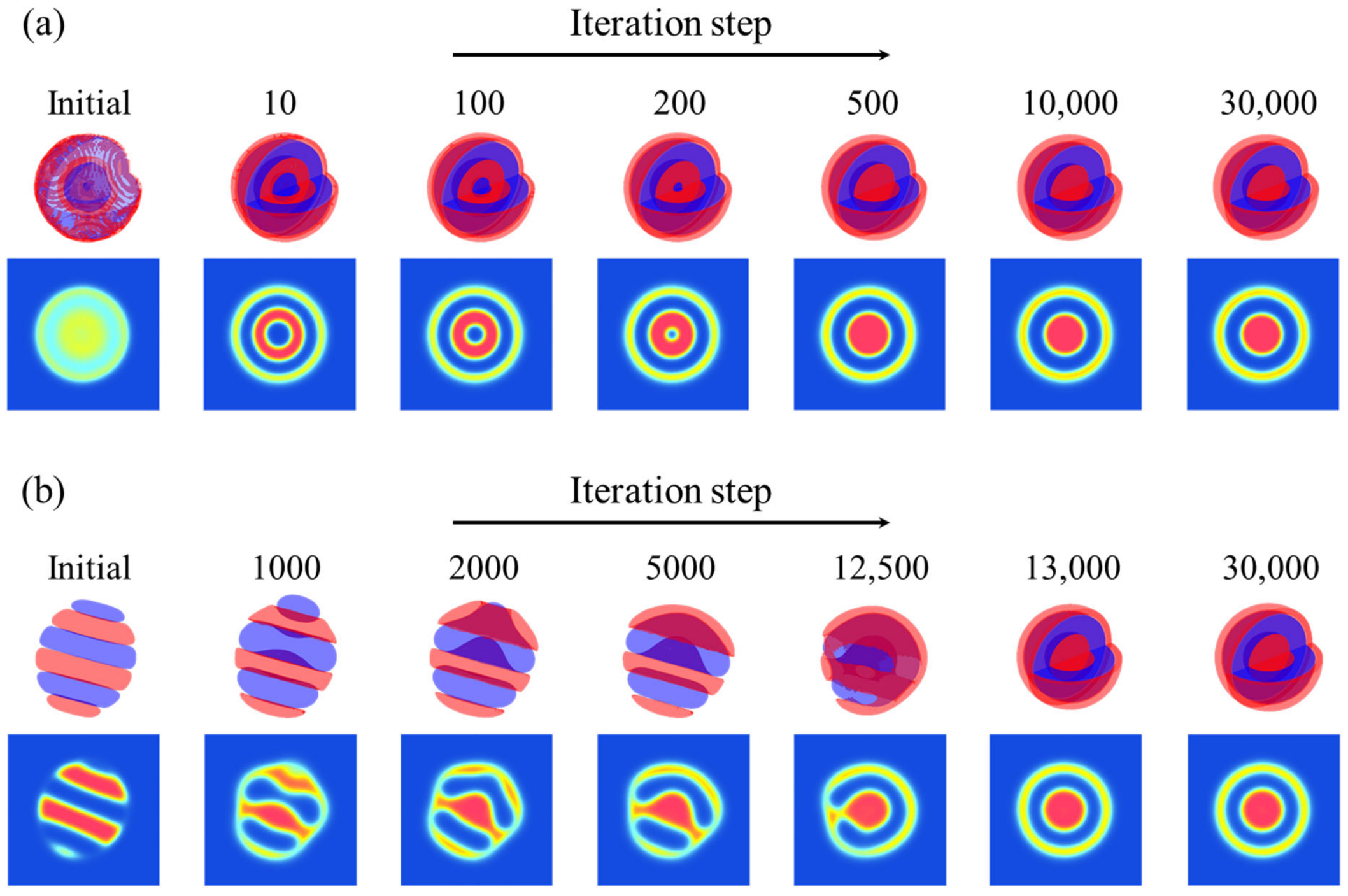}}
\caption{
\textsf{The inner structural evolution during the formation of onion-like particles. (a)~A weak-ordered particle changes to an onion-like one upon varying $N\chi_{\rm{AB}}$ from $8$ to $27$. (b)~A striped ellipsoidal particle changes to an onion-like particle upon varying $N\chi_{\rm{AB}}$ from $27$ to $19$. In both (a) and (b) the solvent is selective, $N\chi_{\rm{AC}}=21$ and $N\chi_{\rm{BC}}=9$. The first and third rows are isosurface plots, where A-rich domains are colored in blue and B-rich domains are in red. The second and fourth rows are cross-section contour plots of the B component indicated by the red color.
}}
\label{fig5}
\end{figure*}
%%%%%%%%%%%%%%%%%%%%%%%%%%%%%%%%%%%%%%%%

We focus now on steps ($\it{iv}$) and ($\it{v}$) of strategy I, which correspond to the formation of onion-like particles. Figure~\ref{fig5} shows the structural evolution of particle during the two transition processes, which are (a) weak-ordered to onion-like; and, (b) striped ellipsoid to onion-like. In both figure parts, the top row shows isosurface plots and the bottom row shows cross-section contour plots of the B component (the B domains are marked in red).  

Figure~\ref{fig5}a shows a cooling process starting with a weak-ordered particle where $N\chi_{\rm{AB}}$ varies from $8$ to $27$. The particle is immersed in a selective solvent, where $N\chi_{\rm{AC}}=21$ and $N\chi_{\rm{BC}}=9$ indicate that the solvent prefers the B component. The structural evolution in figure~\ref{fig5}a has two stages: (a) the solvent preference induces microphase separation (before iteration step $200$), and (b) $N\chi_{\rm{AB}}$ induces microphase separation (after iteration step $200$). For iteration steps $=10$, $100$ and $200$ of the first stage, the cross-section contour plots (the second row of figure~\ref{fig5}a) show that the BCPs inside the particle gradually separate into multiple layers. The characteristic length of these layers is quite different from the corresponding periodicity of the bulk lamellar phase. Therefore, this microphase separation is mainly induced by the solvent preference.

As the iterations continue, the number of layers gradually decreases, entering into stage (b). This second stage starts from around the $500$th iteration step, at which onion-like structure with two B layers and one A layer is formed. It is nearly unchanged throughout iteration step 10,000 and even 30,000. This indicates that in stage (b), the particle converges into an equilibrium state that is thermodynamically determined by $N\chi_{\rm{AB}}$.

Onion-like particles can also be formed by heating an ellipsoidal particle. In figure~\ref{fig5}b, a striped ellipsoidal particle is heated by varying $N\chi_{\rm{AB}}$ from $27$ to $19$ in the same selective solvent used in figure~\ref{fig5}a. The structural evolution in figure~\ref{fig5}b shows that the lamellae first curve, as shown by the isosurface plots and cross-section contour plots at iteration step 1,000 and 2,000. Then, one component surrounds the other (see iteration step 5,000 and 12,500), leading to the formation of onion-like particles. This transition is in good agreement with the experimental results of Avalos et al~\cite{Avalos18}. Using thermal annealing to investigate the morphological transition of PS-b-PI particles, they obtained a transition from a striped ellipsoid to onion-like particle as temperature was increased.

%fig6
%%%%%%%%%%%%%%%%%%%%%%%%%%%%%%%%%%%%%%%%
\begin{figure*}[h!t]
{\includegraphics[width=0.5\textwidth,draft=false]{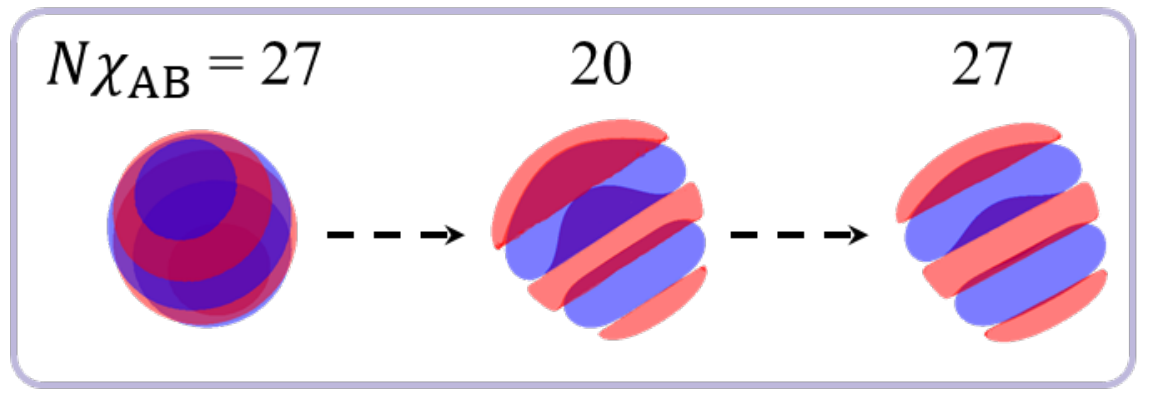}}
\caption{
\textsf{An example for the choice of $N\chi_{\rm{AB}}$ in step ($\it{iv}$) (e.g., see figure~\ref{fig1}) for the transition from an ellipsoidal to onion-like particle. In a selective solvent environment, heating the ellipsoidal particle from $N\chi_{\rm{AB}}=27$ to $20$, and then cooling it to $N\chi_{\rm{AB}}=27$ results in a deformed ellipsoidal particle.
}}
\label{fig6}
\end{figure*}
%%%%%%%%%%%%%%%%%%%%%%%%%%%%%%%%%%%%%%%%

Similar to the formation of striped ellipsoidal particle, the choice of $N\chi_{\rm{AB}}$ in the intermediate step ($\it{iv}$) is also important for the formation of onion-like particle. Our calculations show that $N\chi_{\rm{AB}}$ has to be less than $19$ ($N\chi_{\rm{AB}} \le 19$) in step ($\it{iv}$). Otherwise, onion-like particle cannot be obtained. For example, figure~\ref{fig6} shows that a deformed ellipsoidal particle (the middle isosurface plot) is formed after heating a striped ellipsoidal particle from $N\chi_{\rm{AB}}=27$ to an intermediate value of $20$. Such a deformed ellipsoidal particle cannot further evolve into onion-like particle by applying a further cooling (varying $N\chi_{\rm{AB}}$ from $20$ to $27$), as shown in the second step of figure~\ref{fig6}.

%%%%%%%%%%%%%%%%%%%%%%%%%%%%%%%%%%%%%%%%%%%%%%%%%%
\section{Conclusions}
%%%%%%%%%%%%%%%%%%%%%%%%%%%%%%%%%%%%%%%%%%%%%%%%%%

%fig7
%%%%%%%%%%%%%%%%%%%%%%%%%%%%%%%%%%%%%%%%
\begin{figure*}[h!t]
{\includegraphics[width=0.75\textwidth,draft=false]{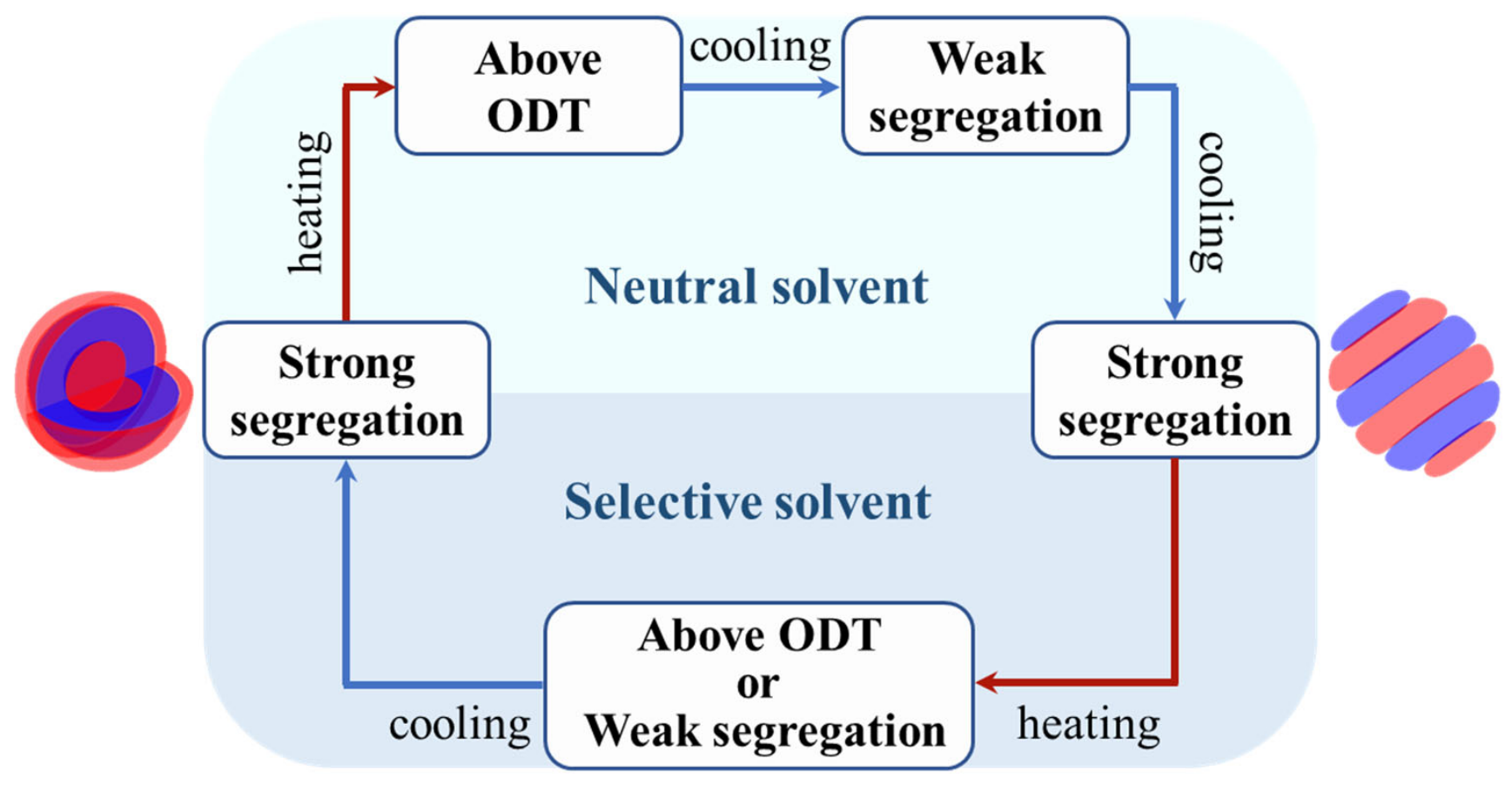}}
\caption{
\textsf{Schematic illustration of the reversible shape transition between onion-like particle and striped ellipsoidal particle. Employing heating followed by a two-step cooling in a neutral solvent environment leads to the transition from an onion-like to striped ellipsoidal particle. While the reverse transition from striped ellipsoidal to onion-like particle can be achieved via heating followed by a one-step cooling in selective solvent. Note that in order to achieve such a shape transition, the value of $N\chi_{\rm{AB}}$ has to be within a certain range for each heating or cooling process.
}}
\label{fig7}
\end{figure*}
%%%%%%%%%%%%%%%%%%%%%%%%%%%%%%%%%%%%%%%%

We have explored the kinetic path of the formation of BCP particles using DSCFT. We explain the formation mechanism of striped ellipsoids, onion-like particles and double-spiral lamellar particles. We predict a reversible transition path for particles changing from onion-like to striped ellipsoidal, and then back to onion-like by varying the Flory-Huggins parameter (or equivalently the temperature), as shown schematically in figure~\ref{fig7}. Following such a path, an onion-like particle is first immersed in a neutral solvent environment (no preference to one of the two BCP components), and is heated to a temperature above the ODT. Then, a two-step cooling process is applied to enter the strong-segregation regime, leading to the formation of striped ellipsoidal particles. Subsequently, the striped ellipsoidal particle is transferred into a selective solvent environment, and is heated into the weak-segregation regime or above ODT. Upon further applying a one-step cooling into the strong-segregation regime, an onion-like particle can be obtained again. We also show that the double-spiral lamellar particle can be obtained by adjusting the kinetic path. For example, changing one step in the kinetic path shown in figure~\ref{fig7}, we heat an onion-like particle in selective solvent instead of neutral one, and then apply a two-step cooling in neutral solvent environment resulting in the double-spiral lamellar particle.

Besides the shape transition, we investigate the evolution of the inner structure of a BCP particle during the formation process of striped ellipsoidal and onion-like particle, respectively. We find that an intermediate particle with inner bi-continuous structure is easily formed for BCP particles immersed in a poor solvent. We show that kinetic paths that can destroy such bi-continuous structures and form layered ones are important for the formation of striped ellipsoidal particles. 

Moreover, by investigating the formation process of onion-like particles, it is found that the BCPs inside the particle undergoing two stages of microphase separation. The particle first forms layered structure that is induced by solvent preference. Note that such a structure has a characteristic length that is different from the bulk periodicity of di-BCPs. Then, the layered structure gradually evolves into a lamellar phase that is thermodynamically determined by $N\chi_{\rm{AB}}$. Our calculations highlight the importance of the kinetic path for the formation of ordered and defect-free BCP particles. We hope that our predictions will be verified in future experiments and will have direct implications for industrial applications.

In this paper, our focus is on symmetric diblock copolymers. Asymmetric copolymers, such as BCPs with $f=0.3$, typically form oblate particles when they are immersed in a poor solvent, which corresponds to cylindrical phases in bulk. Generally, particles adjust their shapes to match the commensurability between particle size and BCPs' natural periodicity. In addition to diblock copolymers, other types of BCPs, including triblock copolymers and bottlebrush block copolymers, can form particles with various shapes and inner structures. For example, tulip-bulb particles and particles with interpenetrating networks have been observed~\cite{Steinhaus18,Qiang19}. However, obtaining these complex particles in both experimental and theoretical research is challenging due to their tendency to be trapped in metastable states during the formation process, resulting in the fact that the formation mechanism of these complex particles remains poorly understood. We leave these issues for future studies.

%%%%%%%%%%%%%%%%%%%%%%%%%%%%%%%%%%%%%%%%%%%%%%%%%
\section{Methods Section}
%%%%%%%%%%%%%%%%%%%%%%%%%%%%%%%%%%%%%%%%%%%%%%%%%%%%%%

We employ a three-dimensional (3D) dynamic self-consistent field theory (DSCFT)~\cite{Fraaije97,Huang21} to investigate the kinetic paths of the formation of block copolymer (BCP) particles. DSCFT simulation has been used and also recently developed by Schmid et al to investigate the dynamics of phase separation of inhomogeneous polymer systems~\cite{muller05,Qi17,Mantha20}. Here, we consider a system that is composed of symmetric AB di-BCP chains and C homopolymers acting as a solvent. The symmetric di-BCP has a fraction of the A block $f=0.5$, and forms a lamellar phase in the bulk. The C homopolymer is taken to be a poor solvent for the BCPs. The Flory-Huggins interaction parameters between these three components (A, B, and C) are denoted as $N\chi_{\rm{AB}}$, $N\chi_{\rm{AC}}$, and $N\chi_{\rm{BC}}$. For such a system, the free energy has been discussed in detail elsewhere~\cite{Rubinstein03,Fredrickson06}, and is given by
\begin{equation}
\begin{aligned}
\frac{F}{k_{\rm{B}}T}=&\frac{1}{V}\int \rm{d^3}{\bf r} \left.\big[ \it{N}\chi_{\rm{AB}}\phi_{\rm A}\phi_{\rm B}+N\chi_{\rm{AC}}\phi_{\rm A}\phi_{\rm C}+N\chi_{\rm{BC}}\phi_{\rm B}\phi_{\rm C}-\omega_{\rm A}\phi_{\rm A}-\omega_{\rm B}\phi_{\rm B}-\omega_{\rm C}\phi_{\rm C} 
\right.\\
&\left.+\frac {\kappa}{2}\left(\phi_{\rm A}+\phi_{\rm B}+\phi_{\rm C}-1\right)^{2} \right.\big] -c\ln Q_{\rm {AB}}-\frac{1-c}{\alpha}\ln Q_{\rm C}     
\end{aligned}
\end{equation}
where $\phi_{i}$ is the local density of ${i}$-component (${i}$=$\rm{A}$, $\rm{B}$ or $\rm{C}$), $\omega_{i}$ are the auxiliary fields coupled to $\phi_{i}$, and $Q_{\rm{AB}}$ and $Q_{\rm C}$ are the single-chain partition function for di-BCP and homopolymer, respectively. The average volume fraction of di-BCP is $c$, the chain length ratio between homopolymer and di-BCP is $\alpha$, and the Helfand-type coefficient, $\kappa$, enforces a finite compressibility. 

The temporal evolution of the local density $\phi_{i}({\bf r},t)$ at position $\bf r$ and time $t$ is assumed to be driven by the gradient of chemical potential, $\mu_{i}(\bf r)$, that is calculated from $\mu_{i}(\bf r) \equiv \delta {\it F}[\phi_{\it i}]/\delta \phi_{\it i}$. Then, the evolution of $\phi_{i}({\bf r},t)$ is determined by the diffusion equation~\cite{Ji20}
\begin{equation}
\begin{aligned}
\frac{\partial \phi_{i}({\bf r},t)}{\partial t}=M_{i}\nabla{\cdot}\phi_{i}({\bf r},t) \nabla \mu_{i}(\bf r)+\eta_{\it i}({\bf r},\it t)   
\end{aligned}
\end{equation}
where $M_{i}$ denotes the mobility coefficient of the $i$-type component taken as a constant, and $\eta_{i}({\bf r},\it t)$ is the thermal noise. In each time step, the self-consistent condition of the $\phi_{i}$ and $\omega_{i}$ is imposed by using the Fletcher-Reeves non-linear conjugate-gradient method.

In the calculations, we fix the volume fraction of di-BCP, $c=0.11$, and the ratio of chain length between homopolymer and di-BCP, $\alpha=0.5$, and the Helfand-type coefficient, $\kappa=100$. The 3D calculation box has a size of $L_x \times L_y \times L_z=16R_g \times 16R_g \times 16R_g$, where $R_g$ is the chain radius of gyration and the box is discretized into $64 \times 64 \times 64$ lattice sites. In Eq (2), the discrete time step is set to $\Delta t=0.2\tau$, where $\tau=\Delta x^2/M_i k_B T$ is the simulation time unit. The thermal noise terms, $\eta_i$, are discretized using the scheme proposed by van Vlimmeren et al.~\cite{Vlimmeren96}, and the noise scaling parameter $\Omega$ is chosen as 100. These are the same values as were used in most of the previous works of Fraaiji and co-workers~\cite{Fraaije97,Zvelindovsky98,Vlimmeren99}.

Due to the high computational cost of implementing nonlocal coupling models in DSCFT calculations, we use the Cahn-Hilliard-Cook equations with local coupling approximation. Moreover, we assume same mobility coefficient for different blocks for simplicity.

\bigskip
%%%%%%%%%%%%%%%%%%%%%%%%%%%%%%%%
{\bf Acknowledgements}~~
%%%%%%%%%%%%%%%%%%%%%%%%%%%%%%%%

This work was supported in part by the NSFC-ISF Research Program, jointly funded by the National Natural Science Foundation of China (NSFC) under grant No.~21961142020 and the Israel Science Foundation (ISF) under grant No.~3396/19, NSFC grants No.~21822302 and ISF grant No. 213/19, the Fundamental Research Funds for the Central University under grant No.~YWF-22-K-101. We also acknowledge the support of the High-Performance Computing Center of Beihang University.

\newpage
%%%%%%%%%%%%%%%%%%%%%%%%%%%%%%%%%%%%%%%%%

\clearpage
\vskip 0.5truecm
\centerline{for Table of Contents use only}
\centerline{\bf The process-directed self-assembly of block copolymer particles}
\centerline{\it Yanyan Zhu, Changhang Huang, Liangshun Zhang$^*$, David Andelman$^*$, and Xingkun Man$^*$}

\begin{figure}[h!t]
{\includegraphics[width=0.8\textwidth,draft=false]{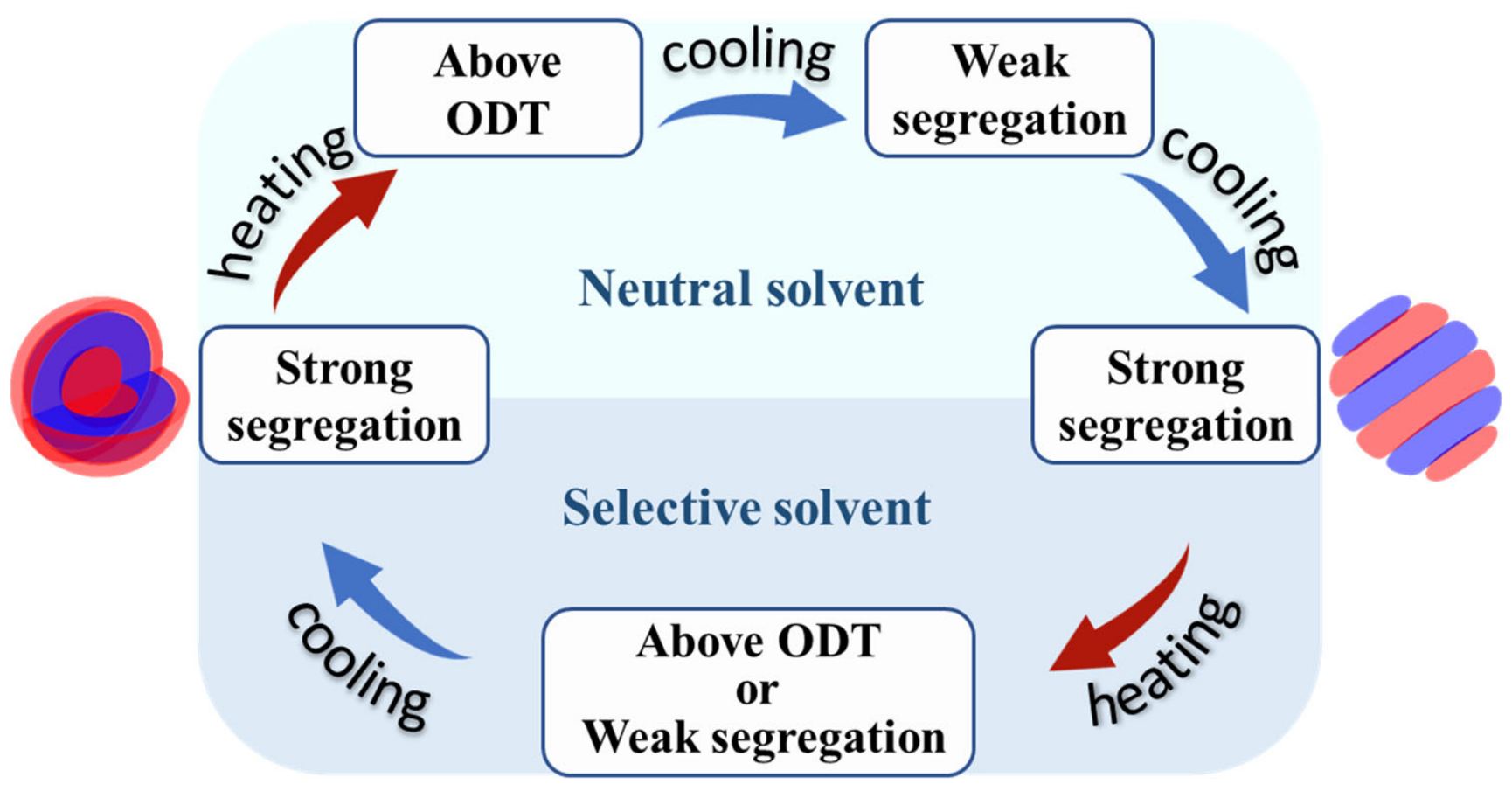}}
\end{figure}
\textsf{The process-directed self-assembly of block copolymer particles leads to the formation of striped ellipsoids, onion-like particles, and double-spiral lamellar particles. According to the dynamic self-consistent field theory, a reversible shape transition between onion-like particles and striped ellipsoidal ones is achieved by controlling the temperature and the selectivity of the solvent towards one of the two BCP components.}

\end{document}